\documentclass[submission,copyright,creativecommons]{eptcs}
\providecommand{\event}{ICLP DC 2021} 
\usepackage{underscore}           
\usepackage{subfig}
\usepackage[ruled,linesnumbered,vlined]{algorithm2e}
\usepackage{pgfplots}
\usepackage{graphicx}
\usepackage{subfig}
\usepackage{todonotes}
\usepackage{amsmath}
\newtheorem{example}{Example}[section]

\title{Compilation of Aggregates in ASP}
\author{Giuseppe Mazzotta\institute{University of Calabria\\Rende,Italy}\email{giuseppe.mazzotta@unical.it}}
\def\titlerunning{Compilation of Aggregates in ASP}
\def\authorrunning{G. Mazzotta}

\begin{document}
\maketitle
\begin{abstract}
Answer Set Programming (ASP) is a well-known problem-solving formalism in computational logic. 
Nowadays, ASP is used in many real world scenarios thanks to ASP solvers. Standard evaluation of ASP programs suffers from an intrinsic limitation, knows as Grounding Bottleneck, due to the grounding of some rules that could fit all the available memory. As a result, there exist instances of real world problems that are untractable using the standard Ground \& Solve approach. In order to tackle this problem, different strategies have been proposed. Among them we focus on a recent approach based on compilation of problematic constraints as \textit{propagators}, which revealed to be very promising but is currently limited to constraints without aggregates. Since aggregates are widely used in ASP, in this paper we extend such an  approach also to constraints containing aggregates. Good results, that proof the effectiveness of the proposed approach, have been reached.
\end{abstract}
\section{Introduction}
Answer Set Programming (ASP)~\cite{DBLP:journals/cacm/BrewkaET11} is a well-known problem-solving formalism in computational logic that is based on the stable model semantics~\cite{DBLP:journals/ngc/GelfondL91}.
ASP systems, such as \textsc{clingo}~\cite{DBLP:conf/iclp/GebserKKOSW16} and \textsc{dlv}~\cite{DBLP:conf/lpnmr/AlvianoCDFLPRVZ17}, made possible the development of many real-world applications.
As a matter of fact, in the recent years, ASP has been widely used for solving problems of game theory~\cite{DBLP:conf/ijcai/AmendolaGLV16}, natural language processing~\cite{DBLP:journals/fuin/Schuller16}, natural language understanding~\cite{DBLP:conf/jelia/CuteriRR19}, robotics~\cite{DBLP:journals/ki/ErdemP18}, scheduling~\cite{DBLP:conf/lpnmr/DodaroM17}, and more~\cite{DBLP:journals/aim/ErdemGL16}. 

A key role in the development of applications is played by system performances. Since ASP system are involved in many real world applications, efficient ASP systems are needed and then the improvement of those systems is an interesting research topic in computational logic.

Traditional ASP systems are based on the ground \& solve approach~\cite{DBLP:journals/aim/KaufmannLPS16}. Stable model computation starts with, a \textit{grounder} module that transforms the input program (containing variables) in its propositional counterpart by substituting variables with possible constants appearing in the program. Afterward, a \textit{solver} module will compute the stable models for the grounded program implementing an extension of the Conflict Driven Clause Learning (CDCL) algorithm~\cite{DBLP:journals/aim/KaufmannLPS16}.

As observed in different contexts~\cite{DBLP:journals/tplp/CuteriDRS17}, the ground \& solve approach suffers from an intrinsic limitation: the combinatorial blow-up of the grounding due to some rules, known as \textit{grounding bottleneck}. In particular, the grounding of these rules could consume all the available memory and then the solver module cannot compute stable models. As a result, there are many problems that are not tractable with the standard approach due to grounding bottleneck.

In order to tackle this problem different approaches have been proposed. One of these strategies is based on the lazy grounding of the rules during model computation. 
In this approach, the grounder and the solver are continuously working together in an iterative way. 
The idea behind lazy grounding is essentially to ground a rule only when its body is satisfied, then the instantiated rules will be added to the solver and then the model computation process restarts. 
Lazy grounding turns out to be very efficient in solving grounding bottleneck problem but obtain bad search performance. 
In order to improve the performance of lazy-grounding approach, the concept of laziness has been relaxed and different strategies have been proposed ~\cite{DBLP:conf/lpnmr/TaupeWF19}.   
Recently, a different solution was proposed in order to solve this issue. The propsed approach is based on the \textit{compilation} of problematic constraints as propagators~\cite{DBLP:journals/tplp/CuteriDRS17,DBLP:journals/tplp/CuteriDRS19}. Basically, Cuteri et al.~\cite{DBLP:journals/tplp/CuteriDRS19} proposed  to translate (or compile) some non-ground constraints into a dedicated C++ procedure in order to skip the grounding of these constraints and simulate them in the solver with this ad-hoc procedure.
This approach reveals to be very promising but, unfortunately, is limited to simple constraints that does not contain aggregates.
Since aggregates are widely used in ASP and their grounding could be very large, we decided to extend the benefits of the compilation based approach also to constraints that contain aggregates.
At early stage of our research, we focused on the compilation of constraints containing count aggregates only and preliminary results were encouraging.
Performances of the solver \textsc{wasp}~\cite{DBLP:conf/lpnmr/AlvianoDLR15} were improved on tested benchmarks.
Then, we decided to push forward this idea and try to extend the compilation also to constraints with sum aggregates and also to simple rules.

\section{Background and Existing Solutions}
\subsection{Answer Set Programming}
An ASP program $\pi$ is a set of rules of the form:
\begin{center}
	$h_1\ |\ldots|\ h_n :- b_1,\ldots,b_m.$
\end{center}
where $n+m>0$, $h_1|\ldots|h_n$ is a disjunction of atoms referred to as \textit{head}, and $b_1,\ldots,b_m$ is a conjunction of literals referred to as \textit{body}. 
If $n=0$, then the rule is called \textit{constraint}, whereas if $m=0$ the rule is called \textit{fact}.

An atom $a$ is an expression of the form $p(t_1,\ldots,t_k)$ where $p$ is a predicate of arity $k$ and $t_1,\ldots,t_k$ are \textit{terms}. A term is an alphanumeric string that could be either a \textit{variable} or a \textit{constant}. According to Prolog notation, if a term starts with a capital letter is a \textit{variable} otherwise is \textit{constant}. If $\forall{i \in \{1,\ldots,k\} }$, $t_i$ is a constant then the atom $a$ is said \textit{ground}.
A \textit{literal} is an atom $a$ or its negation $\sim a$ where $\sim$ denotes the \textit{negation as failure}. Given a literal $l$ it is said \textit{positive} if $l=a$, \textit{negative} if $l=\sim a$.
Given a positive literal $l = a$, we define the \textit{complement}, $\overline{l} = \sim a$, instead, for a negative literal $l = \sim a$, $\overline{l} = \overline{\sim a} = a$. ASP supports also \textit{aggregate atoms}. An aggregate atom is of the form $f (S) \succ T$, where $f (S)$ is an aggregate function, $\succ \in \{=, <,\leq, >,\geq\}$ is a predefined comparison operator, and T is a term referred to as guard. An aggregate function is of the form $f (S)$,
where $S$ is a set term and $f \in \{\#count, \#sum\}$ is an aggregate function symbol. A set term $S$ is a pair that is either a symbolic set or a ground set. A symbolic set is a pair $\{\mathit{Terms} : \mathit{Conj}\}$, where \textit{Terms} is a list of variables and \textit{Conj} is a conjunction of \textit{standard} atoms, that is, \textit{Conj} does not contain aggregate atoms.
A ground set, instead, is a set of pairs of the form $(\overline{t} : \mathit{conj})$, where $\overline{t}$ is a list of constants and \textit{conj} is a conjunction of ground atoms.
Given a program $\pi$ we define the positive dependency graph $G_{\pi} = <V,E>$ as a directed graph where $V=\{p : p $ is a predicate term appearing in $\pi\}$ and $E=\{(u,v) : u \in V, v \in V$ and $\exists{r \in \pi}$ s.t. $\exists$ a positive literal $l$ and an atom $a$ s.t. $l$ is of the form $u(t_1,...,t_k)$ and $l$ appears in the body of $r$, and $a$ is of the form $v(t_1,...,t_n)$ and $a$ appears in the head of $r\}$. $\pi$ is said to be recursive if $G_{\pi}$ is a cyclic graph.
Given a program $\pi$, we define $U_{\pi}$, the \textit{Herbrand Universe}, as the set of all constants appearing in $\pi$ and $B_{\pi}$, the \textit{Herbrand Base}, as the set of all possible ground atoms that can be built using predicate in $\pi$ and constants in $U_{\pi}$. $\mathcal{B}$ denotes $B_{\pi} \cup \overline{B_{\pi}}$. Given a rule $r$ and the Herbrand Universe $U_{\pi}$, we define $ground(r)$ as the set of all possible instantiations of $r$ that can be built by assigning variables in $r$ to constant in $U_{\pi}$. Given a program $\pi$, instead, $ground(\pi)=\bigcup_{r \in \pi} ground(r)$. 
An interpretation $I$ is a set of literals. In particular, $I$ is total if $\forall{a \in B_{\pi}} (a \in I \vee \sim a \in I) \wedge \,\, (a \in I \rightarrow \sim a \notin I)$.  A literal $l$ is true w.r.t $I$ if $l \in I$, otherwise it is false. A ground conjunction \textit{conj} of atoms is true w.r.t $I$ if all atoms in \textit{conj} are true, otherwise, if at least one atom is false then \textit{conj} is false w.r.t. $I$. Let $I(S)$ denote the multiset $[ t_1 | (t_1 ,\ldots , t_n) : \mathit{conj} \in S \wedge \mathit{conj} $ \, is true w.r.t. $I ]$. The evaluation $I(f (S))$ of an aggregate
function $f (S)$ w.r.t. I is the result of the application of $f$ on $I(S)$. 

\begin{example}
Let $A$ be an aggregate atom $A=\#count\{(1:p(1,1)) , (2:p(2,1)),(3:p(3,1))\}>1$ and let $I=\{p(1,1),p(2,1),\sim p(3,1)\}$. $I(S)=[1,2]$, $I(f(S)) = 2 $ since $f=\#count$ so the aggregate atom $A$ is true w.r.t. $I$.
\end{example}
An interpretation $I$ is a \textit{model} for $\pi$ if $\forall{\,r\in ground(\pi) (\forall{\,l \in body(r), l \in I}) \rightarrow (\exists{a \in head(r)}: a \in I})$. The \textit{FLP-reduct} of $\pi$, denoted by $\pi^I$, is the set of rules obtained from $\pi$ by deleting those rules whose body is false w.r.t $I$. 
Let $I$ be a model for $\pi$, $I$ is a \textit{stable model} for $\pi$ if there is no $I' \subset I$ such that $I'$ is a model for $\pi^I$. Given a program $\pi$, $\pi$ is \textit{coherent} if it admits at least one stable model otherwise is \textit{incoherent}.

\subsection{Existing Literature}
In order to solve the grounding bottleneck problem, several attempts have been made ~\cite{DBLP:conf/ijcai/GebserLMPRS18}, including language extensions (such as  constraint programming~\cite{DBLP:journals/tplp/OstrowskiS12,DBLP:journals/tplp/BalducciniL17}, difference logic~\cite{DBLP:conf/iclp/GebserKKOSW16,DBLP:conf/iclp/SusmanL16}) and \textit{lazy grounding} techniques~\cite{DBLP:journals/fuin/PaluDPR09,DBLP:conf/lpnmr/LefevreN09a,Weinzierl2017}. Hybrid formalisms are efficiently evaluated by coupling an ASP system with a solver for the other theory, thus circumventing the grounding bottleneck.
Lazy grounding implementations instantiate a rule only when its body is satisfied to prevent the grounding of rules which are unnecessary during the search of an answer set.
Albeit lazy grounding techniques obtained good preliminary results, their performance is still not competitive with state-of-the-art systems~\cite{DBLP:conf/ijcai/GebserLMPRS18}.
Lazy grounding has been also extended to support aggregates~\cite{DBLP:conf/aaai/BomansonJW19}. To the best of my knowledge, this normalization strategies is limited to monotone aggregates with a lower bound. This approach turns out to be very promising outperfoming the ground \& and solve system \textsc{Clingo} on benchmarks were the grounding is really hard.

Another existing approach is proposed by Cuteri et al.~\cite{DBLP:journals/tplp/CuteriDRS17,DBLP:journals/tplp/CuteriDRS19} and is based on the compilation of constraints into a dedicated procedure, called \textit{propagator}, that supports the solver in the model computation process. There exists two different implementations of this approach, namely \textit{lazy} and \textit{eager}. In more details, given an ASP program $P$ and a set of constraint $C$ such that $C \subseteq P$, if the program $P \setminus C$ is unsatisfiable then the original program is also unsatisfiable.
Otherwise, if $P \setminus C$ admits at least one model $I$ and constraints in $C$ are satisfied w.r.t. $I$, then $I$ is a model for the original program.
Starting from this description the \textit{lazy} implementation build a procedure that automatically understands if a candidate model $I$ satisfies constraints in $C$. If some of these constraints are not satisfied then they will be lazily instantiated in the solver and the model computations starts again, otherwise the process stops returning the model $I$. The \textit{eager} implementation is based on a different approach.
Indeed, in this case the propagator procedure is completely involved in the model computation process. 
Every time that the solver assign a truth value to a literal, either true or false, the propagator is notified and it simulates the instantiation of the constraints in $C$ without storing it in memory, and if it is possible it makes some inferences on the truth values of the literals that have not been assigned yet, in order to prevent constraints failure. Results obtained with this compilation-based approach are very promising but currently the eager approach supports only simple constraints without aggregates.
\begin{figure}[!htb]
   \begin{minipage}{0.48\textwidth}
     \centering
     \includegraphics[width=.7\linewidth]{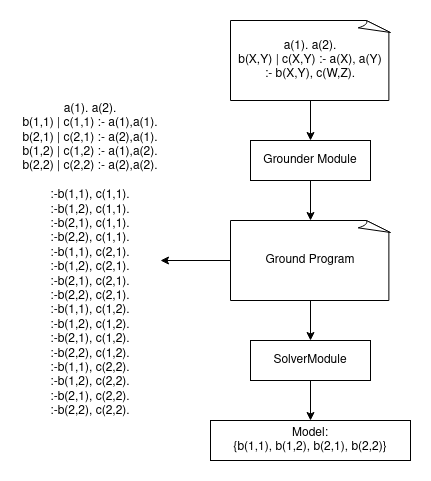}
     \caption{Solving using Ground \& Solve approach}\label{Fig:Data1}
   \end{minipage}\hfill
   \begin{minipage}{0.48\textwidth}
     \centering
     \includegraphics[width=.7\linewidth]{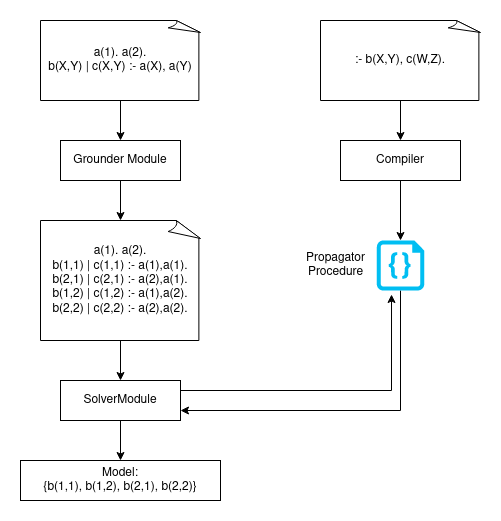}
     \caption{Solving using compilation based approach}\label{Fig:Data2}
   \end{minipage}
\end{figure}
\section{Research Goal}
Approaches based on compilation of constraints revealed to be very promising, outperforming traditional systems in many comparisons. 
However, a significant number of problems, especially hard combinatorial problems from ASP competitions feature aggregates that are not yet supported by compilation-based approaches.
Aggregates are among the standardized knowledge modeling constructs that make ASP effective in representing complex problems. 
So, the extension of the existing compilation approach is an interesting research topic in order to extend the benefits of the compilation to all those ASP programs that cannot be solved by traditional approached.
For this reason, we decided to tackle this problem starting from the compilation of constraints containing aggregates. In particular, a first attempt has been proposed in the paper~\cite{DBLP:conf/cilc/MazzottaCDR20}.
In particular, we presented an extension of the eager compilation for constraints containing \textit{\#count} aggregates and promising results have been obtained.
The next natural step is to extend the compilation approach also to \textit{\#sum} aggregate. Moreover, we also consider to extend the compilation approach also to rules, whereas now it is limited to constraints.

Therefore, our research goals are as follows:
\begin{itemize}
    \item Extend the eager compilation approach also to \textit{\#sum} aggregates.
	\item Support the eager compilation of rules without recursion. 
	\item Support the eager compilation of simple programs containing recursive rules. 
	\item Support, also, the compilation of disjunctive and choice rules in order to provide a complete compilation of ASP program, providing a novel approach to compute stable models of an ASP program.  
\end{itemize}
In principle it can be also combined with the lazy normalization~\cite{DBLP:conf/aaai/BomansonJW19}. However this would not be a trivial combination and might be subject of future research once the compilation approach will be extended to a larger class of programs.

\section{Current Research Status}
Up to now we have extended the compilation to constraints containing \textit{\#count} aggregates. Moreover, we currently have a preliminary implementation of a technique that is able to compile constraints containing also \textit{\#sum} aggregates, whose performance is still not satisfactory. It turns out that compiling \textit{\#sum} aggregates requires several additional optimization techniques and dedicated data structures.

In order to show the working principles of our approach in presence of \textit{\#count} aggregates, we present the following example.

\begin{example}
Let $:- a(X,Z), c(Z), \#count\{Y:b(X,Y)\} >= 2$ be a constraint, then there are several possible propagation steps that can be done. Note that a propagation step is done to avoid possible violations of the constraint.
\begin{itemize}
	\item \textbf{Aggregate propagation.} Let $I = \{a(1,1),\ c(1)\}$ be an interpretation and assume that the solver assigns $b(1,1)$ to true. Starting from $b(1,1)$ it is possible to propagate all the undefined values of the form $b(1,\_)$ to false, since if $b(1,2)$ becomes true the count returns 2 that is greater than or equal to 2 and thus the constraint is violated.
	\item \textbf{Body literal propagation.} Let $I = \{b(1,1),\ b(1,2)\}$ be an interpretation and assume that the solver assigns $a(1,1)$ to true. Note that $I$ satisfies the aggregate, therefore starting from $a(1,1)$ it is possible to propagate $c(1)$ to false because if $c(1)$ becomes true then the constraint is violated.
\end{itemize}
In general, the propagator starts building all possible instantiations of the constraint and looks for undefined values that could be propagated.
If there is a constraint instantiation such that each body literal is true, then we have to check if the count has reached the aggregate's guard minus one.
In this case, we can propagate the aggregate body to false. Note that in this simple case we can just propagate undefined values of $b$ but if the aggregate body is more complex propagation is not so simple. Since literals in the aggregate are in conjunction, in order to propagate a conjunction as false we need only one literal false and then for each possible conjunction we can propagate the last undefined literal of that conjunction.
On the other hand, if there is a constraint instantiation such that exactly one literal is undefined and the aggregate is true then the undefined literal could be propagated as false. So, the propagator procedure is a complex and custom procedure that stores partial interpretation and implements optimized join techniques in order to build constraint instantiation.
\end{example}

In Algorithm~\ref{alg:compileConstraintWithAggregate}, we report a simplified pseudo code propagator for the constraint described in the example. In particular, here we focus only on one propagation case. The complete propagator is based on several procedures, one for every literal in the body of the constraint or in the aggregate body, whose algorithms are similar to Algorithm~\ref{alg:compileConstraintWithAggregate}. 
\paragraph{Notation}
In ordered to better understand Algorithm~\ref{alg:compileConstraintWithAggregate} lets introduce some utility functions. Let $l$ be a ground literal of the form $a(1,2)$ we define:
\begin{itemize}
    \item \textit{getPredicateName()} returns the predicate name of \textit{l}, e.g., "a"
    \item \textit{getTermAt(integer i)} return the i-th term of \textit{l}, e.g., \textit{l}.getTermAt(0) return 1
\end{itemize}

\begin{algorithm}
	\SetKwInOut{Input}{Input}
	\SetKwInOut{Output}{Output}
	\Input{A literal $l$ of the form $a(X,Z)$.}
	\Output{List of propagated literals.}
	\If{l.getPredicateName() == "a"}{
		X = l.getTermAt(0)\\
		Z = l.getTermAt(1)\\
		tupleU = $\bot$\\
		tuplesC = trueLiteralsC.getValuesMatching({Z})\\
		tuplesUC = []\\
		\If{tupleU == $\bot$}{
			tuplesUC = undefinedLiteralsC.getValuesMatching({Z})
		}
		\ForEach{tupleC $\in$ tuplesC $\cup$ tuplesUC}{
			\If{tupleC $\in$ tuplesUC}{
				tupleU = tupleC
			}
			\If{tupleU == $\bot$}{
				tuplesB = trueLiteralsB.getValuesMatching({X})\\
				\If{tuplesB.size() $\geq$ 2}{
					conflict detected on propagator\\
				}\ElseIf{tuplesB.size() == 1}{
					\ForEach{undefinedb $\in$ undefinedLiteralsB.getValuesMatching({X})}{
						propagated undefinedb as False\\
					}
				}
			}\Else{
				tuplesB = trueLiteralsB.getValuesMatching({X})\\
				\If{tuplesB.size() $\geq$ 2}{
					propagate tupleU as False\\	
				}
			}	
		}
	}
	\caption{Example of propagator for one literal.}
	\label{alg:compileConstraintWithAggregate}
\end{algorithm}
The goal of the proposed procedure is to simulate the grounding of the constraint after a literal becomes true. In particular, assume that $a(1,2)$ becomes true, then the procedure is invoked passing $a(1,2)$. So X will be assigned to 1 and Z to 2 (lines 2 and 3, respectively). Afterward, true and undefined values of $c$ matching Z are searched. If the procedure finds a true value of $c$ matching Z then it evaluates the aggregate propagation. If the aggregate is also true then a conflict is found. Otherwise, if the aggregate's guard -1 has been reached the procedure makes a propagation to ensure that the aggregate is made false. On the other hand, if the procedure finds an undefined value of $c$, then it can be propagated as false only when the aggregates is true.

\section{Preliminary Result}

\paragraph{Implementation.} 
We started from the baseline system presented in~\cite{DBLP:conf/ijcai/CuteriDRS20}. The approach has been extended to support the compilation of the propagation of aggregates.
In particular, the compiler has been implemented in C++, and its output, that is the propagator procedure itself, is also C++ code compliant to the \textsc{wasp} propagator interface, and is loaded in the ASP solver as a C++ dynamic library. The source code is not available yet, since we are in a development stage and we are working on a more robust and stable version. 

\paragraph{Experimental Settings.}
We carried our an experimental evaluation to empirically assessed the performance gain of the proposed approach w.r.t. the base solver \textsc{wasp}. Namely, we considered two hard benchmarks of the ASP competitions~\cite{DBLP:journals/ai/CalimeriGMR16}, namely \emph{Combined Configuration} and \emph{Abstract Dialectical Frameworks}, featuring some constraints containing \textit{\#count} aggregates. 

In \emph{Combined Configuration}, the problem is to configure an artifact by combining several components in order to achieve some goals; whereas in \emph{Abstract Dialectical Frameworks} the problem is to find all statements which are necessarily accepted or rejected in a given abstract argumentation framework. In both benchmarks we compile all constraints with aggregates supported by our implementation (i.e., constraints with exactly one \textit{\#count} aggregate). The experiments were run on an Intel Xeon CPU E7-8880 v4 \@ 2.20GHz, time and memory were limited to 10 minutes and 4 GB, respectively.

\paragraph{Results.}
The results are presented in Figure \ref{fig:comb} and Figure \ref{fig:abs} 
as two cactus plots. Overall, our approach is able to boost the performance of  \textsc{wasp}, with the result of obtaining smaller execution times, on average, and more solved instances (3 more instances for \emph{Combined Configuration} and 7 more for \emph{Abstract Dialectical Frameworks}).
The results are very promising, also considering the fact that the benchmarks in the ASP competitions are more oriented towards the evaluation of solving techniques.

\begin{center}
	\begin{figure}[t!]
		\centering
		
		\subfloat[Combined configuration.]{
			\centering
			\begin{tikzpicture}[scale=0.8]
			\pgfkeys{%
				/pgf/number format/set thousands separator = {}}
			\begin{axis}[
			scale only axis
			, font=\Large
			, x label style = {at={(axis description cs:0.5,-0.1)}}
			, y label style = {at={(axis description cs:-0.06,0.5)}}
			, xlabel={Number of instances}
			, ylabel={Execution time (s)}
			, xmin=0, xmax=60
			, ymin=0, ymax=600
			, legend style={at={(0.3,0.96)},anchor=north, draw=none,fill=none}
			, legend columns=1
			, width=0.8\textwidth
			, height=0.4\textwidth
			, ytick={0,120,240,360,480,600}
			, xtick={0,10,20,30,40,50,60}
			, major tick length=2pt
			]

			\addplot [mark size=2pt, color=blue, mark=square] [unbounded coords=jump] table[col sep=semicolon, y index=1] {./combinedConf.csv}; 
			\addlegendentry{\textsc{wasp}}
			
			\addplot [mark size=2pt, color=black, mark=x] [unbounded coords=jump] table[col sep=semicolon, y index=2] {./combinedConf.csv}; 
			\addlegendentry{\textsc{wasp-eager-aggr}}
			\end{axis}
			
			\end{tikzpicture}\label{fig:comb}
		}

		\subfloat[Abstract dialectical frameworks.]{
			\centering
			\begin{tikzpicture}[scale=0.8]
			\pgfkeys{%
				/pgf/number format/set thousands separator = {}}
			\begin{axis}[
			scale only axis
			, font=\Large
			, x label style = {at={(axis description cs:0.5,-0.1)}}
			, y label style = {at={(axis description cs:-0.06,0.5)}}
			, xlabel={Number of instances}
			, ylabel={Execution time (s)}
			, xmin=0, xmax=120
			, ymin=0, ymax=600
			, legend style={at={(0.3,0.96)},anchor=north, draw=none,fill=none}
			, legend columns=1
			, width=0.8\textwidth
			, height=0.4\textwidth
			, ytick={0,120,240,360,480,600}
			, xtick={0,20,40,60,80,100,120}
			, major tick length=2pt
			]

			\addplot [mark size=2pt, color=blue, mark=square] [unbounded coords=jump] table[col sep=semicolon, y index=1] {./abstractDialectical.csv}; 
			\addlegendentry{\textsc{wasp}}
			
			\addplot [mark size=2pt, color=black, mark=x] [unbounded coords=jump] table[col sep=semicolon, y index=2] {./abstractDialectical.csv}; 
			\addlegendentry{\textsc{wasp-eager-aggr}}
			\end{axis}
			
			\end{tikzpicture}\label{fig:abs}
		}
		\caption{Preliminary experimental results on two hard benchmarks taken from ASP competition.}
	\end{figure}
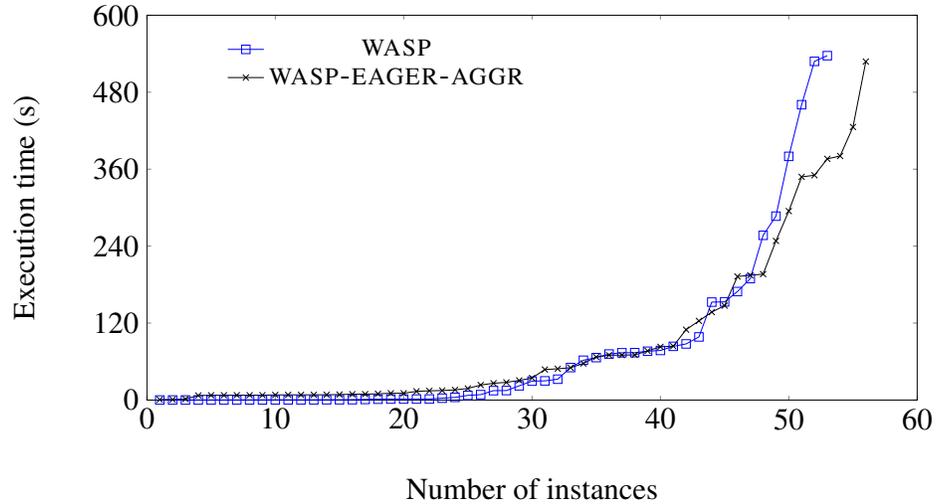
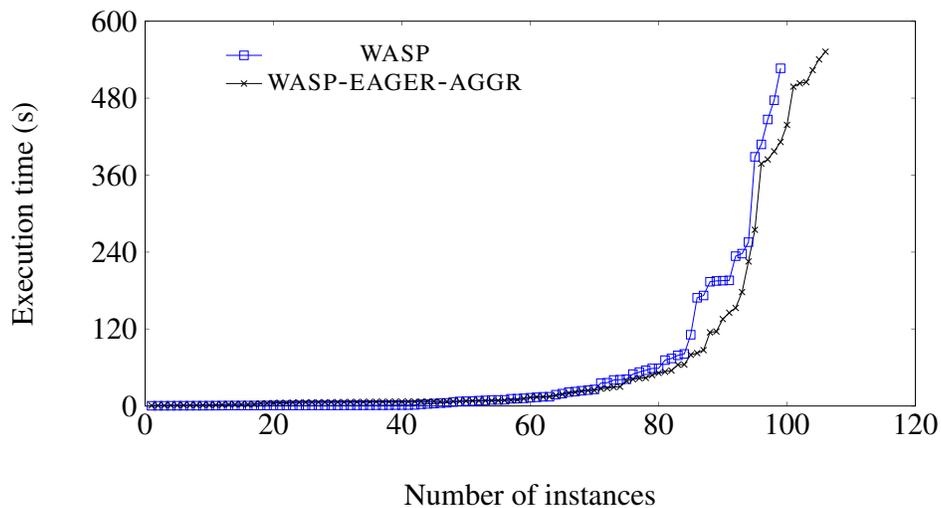
\end{center}

\section{Conclusion}
In this paper we provided an overview of new strategies to deal with grounding bottleneck in presence of aggregates.
Our research is currently focused on compilation based approach, in particular, on eager compilation of problematic constraints possibly containing \textit{\#count} aggregates. Obtained results were very promising and the extension of this approach to cover other  aggregates and rules might lead to an improved of the performance of existing ASP solvers. In particular, we are currently able to compile only simple constraints and also constraint with aggregates but our goal is to move towards the compilation of general rules.
A first attempt has been made together with a rewriter that uses rules to rewrite aggregates, but we are in development stage and no experimental analysis has been conducted yet. The crucial point of the overall work is how to generate procedures that have to be smart enough to save as much space as possible, in order to avoid grounding bottleneck, but at the same time they have to be also time efficient.
We plan to extend the experimental analysis in order to include different benchmarks and to ensure that our approach is able to solve programs that are currently not solvable by state-of-the-art solvers.
Moreover, we also plan to evaluate the impact of our techniques also on problems that are not affected by grounding bottleneck, in order to estimate the overhead of our approach also in this case.
Such analysis might lead to a better understanding of the solving procedure and it can lead to the implementation of (automated) systems that are able to reason, in some way, about the sub-program to compile and afterward create a procedure that is strictly customized on that sub-program.

\bibliographystyle{eptcs}
\bibliography{refs}
\end{document}